\documentclass[conference]{IEEEtran}

\makeatletter
\def\endthebibliography{%
	\def\@noitemerr{\@latex@warning{Empty `thebibliography' environment}}%
	\endlist
}
\makeatother

\IEEEoverridecommandlockouts
\usepackage{cite}
\usepackage{amsmath,amssymb,amsfonts}
\usepackage{algorithmic}
\usepackage{graphicx}
\usepackage{textcomp}
\usepackage{xcolor}
\usepackage{subfigure}
\usepackage{mathtools}

\usepackage{tikz}
\usetikzlibrary{shapes.arrows}
\usetikzlibrary{plotmarks}
\usetikzlibrary{patterns}
\usetikzlibrary{shapes,shapes.geometric,arrows,fit,calc,positioning,automata}
\usetikzlibrary{arrows}
\usetikzlibrary{shapes.multipart}
\usetikzlibrary{decorations.pathreplacing}
\usetikzlibrary{patterns}
\usepackage{adjustbox}
\usepackage{gincltex}

\def\BibTeX{{\rm B\kern-.05em{\sc i\kern-.025em b}\kern-.08em
		T\kern-.1667em\lower.7ex\hbox{E}\kern-.125emX}}

\begin{document}

\title{Age of Information in Multi-hop Networks \\
with Priorities
}

\author{\IEEEauthorblockN{Olga Vikhrova\IEEEauthorrefmark{1}, Federico Chiariotti\IEEEauthorrefmark{2}, Beatriz Soret\IEEEauthorrefmark{2},
Giuseppe Araniti\IEEEauthorrefmark{1}, Antonella Molinaro\IEEEauthorrefmark{1}, Petar Popovski\IEEEauthorrefmark{2}}
\IEEEauthorblockA{\IEEEauthorrefmark{1}\textit{DIIES Department, University Mediterranea of Reggio Calabria, Italy}, \\
emails: olga.vikhrova@unirc.it, araniti@unirc.it,  antonella.molinaro@unirc.it \\
\IEEEauthorrefmark{2}\textit{Department of Electronic Systems, Aalborg University, Denmark}, emails: fchi@es.aau.dk, bsa@es.aau.dk, petarp@es.aau.dk}
}

\maketitle

\begin{abstract}
Age of Information is a new metric used in real-time status update tracking applications.
It measures at the destination the time elapsed since the generation of the last received packet.
In this paper, we consider the co-existence of critical and non-critical status updates in a two-hop system, for which the network assigns different scheduling priorities. Specifically, the high priority is reserved to the packets that traverse the two nodes, as they experience worse latency performance. We
obtain the distribution of the age and its natural upper bound termed \emph{peak age}.
We provide tight upper and lower bounds for priority updates and the exact expressions for the non-critical flow of packets with a general service distribution. The results give fundamental insights for the design of age-sensitive multi-hop systems.  
\end{abstract}


\section{Introduction}
\label{sec:Introduction}

The \textit{Age of Information} (AoI) \cite{kaul2012real}, \cite{kaul2011minimizingaoi} characterizes the freshness of the information from the receiver's perspective, and it has been proved to be a proper metric in many real-time and context-aware Internet of Things (IoT) applications \cite{abdelmagid2019ontheroleofaoi}. In these applications, the end  receiver is interested in a fresh knowledge of the remotely controlled system,  rather  than  the  packet  delay. Besides the average age, the \textit{Peak Age of Information} (PAoI)  \cite{huang2015optimizing} is a byproduct of the age process that quantifies the worst case.

There are many examples of age-sensitive IoT applications.
In \cite{xu2020optimizing}, the authors consider a Mobile Edge Computing (MEC) system and investigate the impact that pre-processing the raw data collected from sensors has in the age performance. 
Another example is given in \cite{stamatakis2020optimal}, which addresses the problem of the optimal status update generation in a wireless system where the source of updates runs applications with regular IoT traffic and AoI-sensitive traffic.
Finally, the role of satellites in tracking applications for wide-area sensor and vehicular networks is growing due to their natural way to provide ubiquitous coverage for the massive IoT in areas where cellular communications are not available or less cost-effective \cite{bacco2018iottraffic}. As explained in \cite{soret2020latency}, Low Earth Orbit (LEO) satellites organised in a constellation may collect the status updates and forward them over the inter- or intra- satellite links to the ground station.

\begin{figure}[hbt!]
    \centering
    \includegraphics{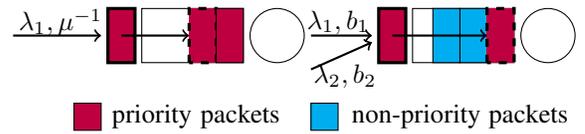}
    \caption{System model as two FCFS queues in tandem with priorities at the second node.}
    \label{fig:SystemModle}
\end{figure}

A close examination of the above-mentioned works reveals the common features of the tracking update systems and existing research gaps. A single queuing system can capture the timeliness of information only between two directly communicating instances, but it fails to give adequate results in multi-hop networks, i.e., when status updates are forwarded over one or several relay nodes. Another element is the existence of heterogeneous requirements and paths: different services should be treated according to their priority level, and status updates might use different entry points to the communication system. 
This motivates us to consider a general multi-hop communication system with traffic arrivals at the intermediate nodes and different priorities for the status updates. For our analysis, we take the illustrative case of two nodes, where  status update packets sent via the relay (the first node) takes priority over the updates sent directly to the monitor (the second node) as shown in Fig.~\ref{fig:SystemModle}. Priority packets preempt all non-priority packets in the queue of the second node but do not impact the ongoing service.
This priority policy will improve the performance of the status updates that need the relay to reach the destination, reducing the difference in performance between the two paths.

In this paper we obtain the distribution of the AoI and the PAoI using the Laplace-Stiltjes Transform (LST) for the system of interest. We also give the distribution of the \textit{System Delay} of priority packets that traverse the two nodes, while the system delay only at one node was known before. Unlike previous works on AoI with packets prioritization, we consider a general service time distribution and more complex system model with relay.
We also give closed-form expressions for the average AoI, PAoI and system delay of non-priority packets and tight bounds for priority packets, while the moments of higher orders can be derived from the given LST expressions.

The rest of the paper is organized as follows. In section \ref{sec:RelatedWorks} we introduce related works on the AoI, and describe the system model in section \ref{sec:SystemModel}. The metrics of interests are given in section \ref{sec:PAoI}, while the numerical results are discussed in section \ref{sec:Results}. The concluding remarks are given in the last section.

\section{Related works}
\label{sec:RelatedWorks}

A system design similar to ours has been considered in \cite{li2020ageoriented}. Authors investigate the average AoI when the status update can be delivered either over the less reliable direct link or over the two-hop relay link with better reliability. However, all packets at the second node have been treated equally. In \cite{xu2020optimizing} only average PAoI is given for the two-hop tandem exponential queues with  multiple sources. Authors in \cite{kuang2019age} study the average AoI of a two-hop system with packet arrivals only at the first node and zero-waiting policy at the second node.

In \cite{inoue2019general} authors derive a general formula for the stationary distribution of the AoI in terms of the system delay and the PAoI for a wide class of $G/G/1$ systems with a single source under the general FCFS and Last Come First Serve (LCFS) packet management policies with various preemption and packet discarding options. However, LCFS policy can not be applied to the systems where packets carry incremental information and can not be discarded.

The idea of assigning different priorities to the update packets has been discussed for the first time in \cite{kaul2018priority}. The average AoI is given for an exponential single-server system with a shared queue and LCFS discipline, where the arrived packet preempts another packet either in service or in waiting only if it has higher priority. In \cite{xu2019peak} authors focus on a queuing system with $k$ classes of priorities, different buffer sizes and queuing disciplines. In particular, the different combinations of infinite queues with FCFS and LCFS disciplines and queues with a single place to wait are considered. The exact expressions of the expected PAoI are given for the general service time distribution if the queues are infinite and for the exponential service time if the queue size is one, while the tight bounds have been calculated for the remaining scenarios. The above-mentioned works with the packet`s prioritization are limited to the single-node systems.

\section{System model}
\label{sec:SystemModel}

We consider a two-hop network with intermediate traffic. Sources generate packets with status updates according to a Poisson process with rate $\lambda$. With probability $p$ priority packets arrive at the first node and with probability $1-p$ all remaining non-priority packets arrive directly to the second node, $\lambda_1 = p \lambda$ and $\lambda_2 = (1-p) \lambda$.
Such a network is modeled as two tandem queues connected in series with packet prioritization in the second queue. 
In particular, both queues apply the general FCFS discipline but in the queue of the second node all packets coming originally from the first node (priority packets) pre-empt in waiting packets coming directly from the source (non-priority packets).
Non-priority packets see the second node as an $M/G/1$ queue with priorities, while priority packets find $M/M/1$ and $M/G/1$ queues connected in series.

Service times at the first node are limited to the exponential distribution for the sake of mathematical tractability, i.e. to ensure that the departure process from the first node is Poisson. Let $b_{1}$ and $b_{2}$ be the mean service times of priority and non-priority packets packets at the second node. The total system utilization equals to the second node utilization $\rho = \rho_1 + \rho_2$, where $\rho_j = \lambda_j b_j$, $j=\{1,2\}$. Utilization of the first node $\rho_{11} = \lambda_1/\mu$, $\mu^{-1}$ is the mean service time at the first node.

Let $j,i$ denote packet $i$ of priority class $j$. Let $t_{j,i}$ and $t'_{j,i}$ be the time instances of packet $j,i$ arrival to the system (generation of a new status at source) and its departure from the system (updating the status at the monitor). Then $Y_{j,i} = t_{j,i} - t_{j,i-1}$ denotes the random variable (RV) of packet $j,i$ \textit{interarrival} time and $T_{j,i} = t'_{j,i} - t_{j,i}$ corresponds to the RV of the packet's \textit{system delay}. The AoI $\Delta_{j,i}$ at time $t>0$ consists of the AoI $Z_{j,i-1}$ immediately after the departure of the packet $j,i-1$ and the time from $t'_{j,i-1}$ to $t$, i.e. $\Delta_{j,i} = Z_{j,i-1} + (t - t'_{j,i-1})$. In general FCFS systems $Z_{j,i}$ equals to the system delay $T_{j,i}$ if all packets are time-stamped on their arrival. Therefore the PAoI $A_{j,i} = t'_{j,i} - t_{j,i-1} = Y_{j,i} + T_{j,i}$.

In the ergodic system ($\rho < 1$), the probability density function (pdf) of the AoI can be defined as $f_{\Delta_j}(x) = \lambda_j(F_{T_j}(x) - F_{A_j}(x)),\ x \geq 0$, where $F_{T_j}(x)$ and $F_{A_j}(x)$ stand for the Probability Distribution Functions (PDFs) of the system delay and PAoI, respectively \cite{inoue2019general}. 
The Laplace-Stiltjes Transform (LST) $\delta_j(s)$ of the AoI distribution therefore yields:
\begin{equation}
\label{eq:AoI}
    \delta_j(s) = \frac{\lambda_j}{s}(\tau_j(s) - \alpha_j(s)),\ s > 0,
\end{equation}
where \resizebox{0.4\linewidth}{!}{$\tau_j(s) = \int\limits_0^{\infty}e^{-sx}d F_{T_j}(x)$} and \resizebox{0.4\linewidth}{!}{$\alpha_j(s) = \int\limits_0^{\infty}e^{-sx}d F_{A_j}(x)$}.

Priority and non-priority packets arrive to the system independently, their interarrival times are exponentially distributed holding the LST $\lambda_j/(\lambda_j + s)$. System delay $T_{j,i}$ depends on the packets interarrival time $Y_{j,i}$ and the system delay $T_{j,i-1}$, it also depends on the arrival and departure processes of packets of another class. The RV $T_{1,i} = T_{11,i} + T_{12,i}$ while $T_{11,i}$ and $T_{12,i}$ are not independent. In the next section we define the PAoI for packet $j,i$ and then obtain the general distribution of $A_{j}$ for both classes of packets, the similar approach is applied for calculation of the total system delay $T_{1}$ of priority packets. 

Let us give the known distributions of the system delays at each node as preliminaries for further analysis. The system delay $T_{11}$ at the first node (M/M/1) is exponentially distributed with parameter $\theta = \mu - \lambda_1$, the corresponding LST $\tau_{11}(s)$ equals to $\theta/(\theta + s)$. The LST of the system delay $T_{12}$ of priority packets and system delay $T_{2}$ of non-priority packets at the second node are given in~\cite[chapter~8.6]{conway1967scheduling}:
\begin{equation}
\label{eq:SystemDelay12}
    \tau_{12}(s) = \frac{s(1-\rho) + \lambda_2(1-\beta_2(s))}{s-\lambda_1+\lambda_1\beta_1(s)}\beta_1(s),
\end{equation}
\begin{equation}
\label{eq:SystemDelay2}
    \tau_{2}(s) =
    \frac{(1-\rho)(s+\lambda_1-\lambda_1\gamma(s))}{s-\lambda_2+\lambda_2\beta_2(s+\lambda_1-\lambda_1\gamma(s))}\beta_2(s),
\end{equation}
where $\beta_1(s)$ and $\beta_2(s)$ are the LSTs of the service time distributions of priority and non-priority packets at the second node, $\gamma(s)$ stands for the LST of the distribution of the interval $G_{1}$, which elapses from the arrival of a priority packet in the empty queue of the second node until the end of continuous service of priority packets arriving afterwards. This interval is known as a busy period generated by a priority packet and its LST $\gamma(s) = \beta_1(s + \lambda_1 - \lambda_1 \gamma(s))$. The busy period $G_2$ starts from the moment when a non-priority packet arrives to the empty node, therefore its LST is $\beta_2(s + \lambda_1 - \lambda_1 \gamma(s))$. For convenience we give the complete list of notations in Table~\ref{tab:Notations}. 
\begin{table}[]
    \centering
    \caption{List of notations}
    \label{tab:Notations}
    \begin{tabular}{p{0.9cm}p{1.9cm}p{4.8cm}}
        \hline
         \multicolumn{2}{l}{\textbf{Notation}} & \textbf{Definition} \\
         \hline
         \multicolumn{2}{l}{\normalsize{$k$}} & Node index \\
         \multicolumn{2}{l}{\normalsize{$(j,i)$}} & Packet $i$ of priority class $j$\\
         \multicolumn{2}{l}{\normalsize{$t_{j,i}$}} & Packet $(j,i)$ arrival time \\
         \multicolumn{2}{l}{\normalsize{$t'_{j,i}$}} & Packet $(j,i)$ departure time \\
         \multicolumn{2}{l}{\normalsize{$\lambda_{j}$}} & Arrival rate for class $j$ \\
         \multicolumn{2}{l}{\normalsize{$b_{j}$}} & Mean service time for class $j$ \\
         \multicolumn{2}{l}{\normalsize{$\rho_{j}$}} & Second node utilization by class $j$ \\
         \multicolumn{2}{l}{\normalsize{$b = \mu^{-1}$}} & Mean service time at the first node \\
         \multicolumn{2}{l}{\normalsize{$\theta$}} & Mean system delay at the first node \\
         \multicolumn{2}{l}{\normalsize{$\rho_{11}$}} & First node utilization \\
         \hline
         \textbf{RV} & \textbf{LST} & \textbf{Definition} \\
         \hline
         \normalsize{$Y_{j,i}$} & & Packets interarrival time \\ 
         \normalsize{$S_{kj,i}$} & \normalsize{$\beta(s)$}, \normalsize{$\beta_{j}(s)$} & Service time of packet $(j,i)$ at node $k$ \\
        \normalsize{$W_{jk,i}$} & \normalsize{$\omega_{jk}(s)$} & Waiting time of packet $(j,i)$ at node $k$\\
         \normalsize{$T_{jk,i}$} & \normalsize{$\tau_{1k}(s)$}, \normalsize{$\tau_{j}(s)$} & System delay of packet $(j,i)$ at node $k$\\
         \normalsize{$D_{j,i}$} & \normalsize{$\eta_{j}(s)$} & Supplementary to PAoI of packet $j,i$ interval as defined in Fig.~\ref{fig:Cases}\\
         \normalsize{$X_{1,i}$} & \normalsize{$\xi_{1}(s)$} & Supplementary to system delay of packet $j,i$ interval as defined in Fig.~\ref{fig:Cases} \\
         \normalsize{$G_{j,i}$} & \normalsize{$\gamma_{j}(s)$} & Busy period generated by a packet $j,i$ \\
         \normalsize{$\tilde{Z}_{j,i}$} & \normalsize{$\tilde{\zeta}_{j,i}(s)$} & Residual time of interval $Z_{j,i}$ \\ 
         \normalsize{$A_{j}$} & \normalsize{$\alpha_{j}(s)$} & PAoI of class $j$ \\
         \normalsize{$\Delta_{j}$} & \normalsize{$\delta_{j}(s)$} & AoI of class $j$\\
    \end{tabular}
\end{table}

\section{Analysis} 
\label{sec:PAoI}

\subsection{Priority packets}
\label{sec:Priority}

When priority packet $i$ arrives to the system it can be queued in both nodes, queued only in one node or go through two nodes without any queuing delay. 
The presence of non-priority packets at the second node hinders the derivation of the PAoI and system delay distributions.  
We assume that packet $i$ finds the second node free of non-priority packets with the probability $1-\rho_2$. 

\begin{figure}[bth]
    \centering
    \includegraphics{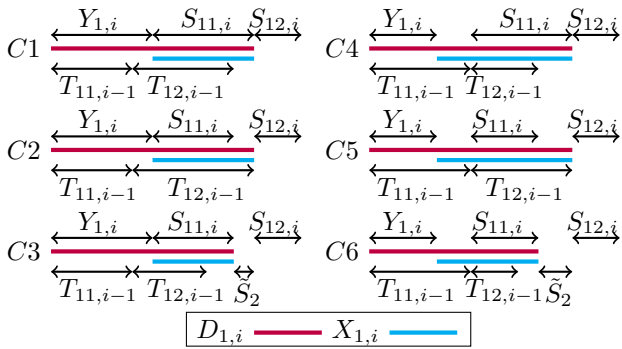}
    \vspace{-0.3cm}
    \caption{PAoI and system delay of priority packets in cases C1 - C6.}
    \label{fig:Cases}
\end{figure}

There are six cases C1--C6 that help to define system delay $T_{1,i}$ and PAoI $A_{1,i}$ of packet $i$ in the system of interest. Let us define intervals $D_{1,i}$ (bold red line) and $X_{1,i}$ (bold blue line) as illustrated in Fig.~\ref{fig:Cases}. 
Let also $\eta_1(s,C_m)$ and $\xi_1(s,C_m)$ denote the LST of the joint distribution of intervals contributing to $D_{1,i}$ and $X_{1,i}$ for a case $C_m$, $m=\{1,\dots,6\}$, respectively.

We define the LST of the system delay $\tau(s,C_m)$ and the PAoI $\alpha(s,C_m)$ for each case. The resulting distributions will be given as a sum of LSTs of the six joint distributions namely $\tau_1(s) = \sum_{m}\tau_{1,i}(s,C_m)$ and $\alpha_1(s) = \sum_{m}\alpha_{1,i}(s,C_m)$.

\begin{itemize}
    \item [C1:] Packet $i$ does not experience any queuing at nodes, therefore the PAoI $A_{1,i} = D_{1,i} + S_{12,i}$ and system delay $T_{1,i} = X_{1,i} + S_{11,i}$. This happens if $T_{11,i-1} < Y_{1,i}$, $T_{12,i-1} + T_{12,i} < Y_{1,i} + S_{11,i}$ and if during the interval $Y_{1,i} + S_{11,i} - T_{11,i-1} - T_{12,i-1}$ all unserved non-priority packets complete their service and no new non-priority packets arrive. Since we assume that packet $i$ finds the second node free of non-priority packets with the probability $1-\rho_2$ and service time $S_{12,i}$ is independent of other intervals, the LST of both metrics can be given as $\alpha(s,C_1) = (1-\rho_2)\eta_1(s,C_1) \beta_1(s)$ and $\tau(s,C_1) = (1-\rho_2) \xi_1(s,C_1) \beta_1(s)$. 
    \item [C2:] Packet $i$ finds the second node busy with packet $i-1$, but its queuing delay at the first node $W_{11,i} = 0$, therefore PAoI $A_{1,i} = D_{i,1} + S_{12,i}$ and system delay $T_{1,i} = X_{1,i} + S_{12,i}$ like in the case C1, but $D_{1,i} = T_{11,i-1} + T_{12,i-1}$, $X_{1,i} = T_{11,i-1} + T_{12,i-1} - Y_{1,i}$. This is true if $T_{11,i-1} < Y_{1,i}$ and $T_{11,i-1} + T_{12,i-1} > Y_{1,i} + S_{11,i}$. The PAoI and system delay distributions in this case give $\alpha(s,C_2) = \eta_1(s,C_2) \beta_1(s)$ and $\tau(s,C_2) = \xi_1(s,C_2) \beta_1(s)$. 
    \item [C3:] Packet $i$ finds the second node busy with a non-priority packet and its waiting time $W_{11,i} = 0$, thus the PAoI $A_{i,1} = D_{1,i} + \tilde{S}_2 + S_{12,i}$ and the system delay $T_{1,i} = X_{1,i} + \tilde{S}_2 + S_{12,i}$, where $\tilde{S}_2$ stands for the LST of the residual service time of a non-priority packet. This happens when $T_{11,i-1} < Y_{1,i}$ , $T_{1,i-1} < Y_{1,i} + S_{11,i}$ like in the case C1, but packet $i$ sees a non-priority packet in service with the probability $\rho_2$. The LST of the PAoI in the case C3 yields $\alpha(s,C_3) = \rho_2 \eta_1(s,C_3) \tilde{\beta}_2(s)\beta_1(s)$, and LST of the $T_{1,i}$ gives $\tau(s,C_3) = \rho_2 \xi_1(s,C_3) \tilde{\beta}_2(s) \beta_1(s)$, where $\tilde{\beta}_2(s) = (1-\beta_2(s))/s\mathbb{E}[S_2]$. 
    \item [C4:] Packet $i$ is queued at the first node, but it finds the second node empty upon the arrival. The PAoI $A_{1,i}$ and system delay $T_{1,i}$ are defined as in the case C1, but in the case C4 $T_{11,i} > Y_{1,i}$ and $T_{12,i} < S_{S_{11,i}}$, in particular $\alpha(s,C_4) = (1-\rho_2)\eta_1(s,C_4) \beta_1(s)$ and $\tau(s,C_4) = (1-\rho_2) \xi_1(s,C_4) \beta_1(s)$.
    \item [C5:] Packet $i$ is delayed by the packet $i-1$ in both nodes, if $T_{11,i} > Y_{1,i}$ and $T_{12,i} > S_{i}$. Given that $A_{1,i} = D_{1,i} + S_{12,i}$ and $T_{1,i} = X_{1,i} + S_{12,i}$ the distribution of PAoI $\alpha(s,C_2) = \eta_1(s,C_2) \beta_1(s)$, the distribution of system delay $\tau(s,C_2) = \xi_1(s,C_2) \beta_1(s)$ in terms of LST.
    \item [C6:] Packet $i$ is queued at the first node and finds the second node busy with a non-priority packet, then like in the case C3 $\alpha(s,C_6) = \rho_2 \eta_1(s,C_6) \tilde{\beta}_2(s)\beta_1(s)$ and $\tau(s,C_6) = \rho_2 \xi_1(s,C_6) \tilde{\beta}_2(s) \beta_1(s)$ given that $T_{11,i} > Y_{1,i}$ and $T_{12,i} < S_{i}$.
\end{itemize}

We now need to calculate the LST of $D_{i,1}$ and $X_{1,i}$ for each case. These intervals are equally defined for the cases C1 and C3, and C4 and C6, therefore we give their derivations with double indexes $\{13\}$ and $\{46\}$.

\paragraph{Cases C1 and C3} we denote the PDF of $D_{1,i}$ as $F_{D_1}(z,C_{13}) =  \mathbb{P}\{D_{1,i} < z, C_{13}\}$. Given that $T_{11,i-1} < Y_{1,i}$ and $T_{1,i-1} < Y_{1,i} + S_{11,i}$ we calculate it as follows:
\begin{align}
\resizebox{0.9\linewidth}{!}{$
    F_{D_1}(z,C_{13}) = \smashoperator{\int\limits_0^{z}} dF_{Y_1}(y) \smashoperator{\int\limits_0^{y}} dF_{T_{11}}(t) \int\limits_0^{z-y} dF_S(x) \smashoperator{\int\limits_0^{x+y-t}} dF_{T_{12}}(u)$} 
\end{align}
The LST $\eta_1(s,C_{13}) =  \int\limits_0^{\infty} e^{-sz} dF_{D_1}(z,C_{13})$ yields:
\begin{align}
    \resizebox{0.9\linewidth}{!}{$
    \eta_1(s,C_{13}) = \frac{\lambda_1}{\lambda_1 + s}\beta_{1}(s)\tau_{12}(\lambda_1 + s) - \rho_{11}\beta^2(s)\tau_{12}(\mu + s).$}
\end{align}
Let $F_{X_1}(z,C_{13}) = \mathbb{P}\{X_{1,i} < z, C_{13}\}$ be the PDF of $X_{1,i}$, it can be calculated as 
\begin{align}
\resizebox{0.9\linewidth}{!}{$
    F_{X_1}(z,C_{13}) = \smashoperator{\int\limits_0^{\infty}}dF_{Y_1}(y) 
    \smashoperator{\int\limits_0^{y}} dF_{T_{11}}(t)
    \smashoperator{\int\limits_0^{z}} dF_S(x) \smashoperator{\int\limits_0^{x+y-t}}dF_{T_{12}}(u), $}
\end{align}
and its LST $\xi_{1}(s,C_{13}) = \int\limits_0^{\infty}e^{-sz}dF_{X_1}(z,C_{13})$ yields:
\begin{align}
\resizebox{0.9\linewidth}{!}{$
    \xi_{1}(s,C_{13}) = \tau_{11}(s)\tau_{12}(\lambda_1) - \rho_{11}\tau_{11}(s)\beta_{1}(s)\tau_{12}(\mu + s).$}
\end{align}

\paragraph{Case C2} the PDF of interval $D_{1,i}$ and its LST $\eta_{1}(s,C_2)$ in the case C2 are given as follows:
\begin{align}
\resizebox{0.9\linewidth}{!}{$
    F_{D_{1,i}}(z,C_2) = \smashoperator{\int\limits_{0}^{z}}dF_{T_{11}}(t) \smashoperator{\int\limits_{0}^{z-t}}dF_{T_{12}}(u) \smashoperator{\int\limits_{t}^{t+u}}dF_{Y_1}(y) \smashoperator{\int\limits_{0}^{t+u-y}}dF_S(z),$}
\end{align}
\begin{align}
    \eta_1(s,C_2) =& (1-\rho_{11})\beta_{1}(s)\tau_{12}(s) - \beta_{1}(s)\tau_{12}(s+\lambda_1) + \nonumber \\
    &+ \rho_{11}\beta_{1}(s)\tau_{12}(\mu + s).
\end{align}
The define the PDF of interval $X_{1,i}$ as
\begin{align}
\resizebox{0.89\linewidth}{!}{$
    F_{X_{1,i}}(z,C_2) = \smashoperator{\int\limits_0^{z}} dF_{T_{11}}(t) \smashoperator{\int\limits_0^{z-t}} dF_{T_{12}}(u) \smashoperator{\int\limits_t^{t+u}} dF_{Y_1}(y) \smashoperator{\int\limits_{0}^{t+u-y}} dF_S(z),$}
\end{align}
while its LST $\xi_1(s,C_2)$ gives
\begin{align}
    \xi_1(s,C_2) =& \frac{\lambda_1}{s - \lambda_1}\tau_{11}(s) \tau(\lambda_1) - \rho_{11}\frac{\theta}{s - \lambda_1}\tau_{12}(s) + \nonumber\\
    &+ \rho_{11}\tau_{11}(s)\tau_{12}(\mu + s).
\end{align}

\paragraph{Cases C4 and C6} we define the PDF $F_{D_1}(z,C_{46})$ and $F_{X_1}(z,C_{46})$ in the cases C4 and C6 as 
\begin{align}
\resizebox{0.89\linewidth}{!}{$
    F_{D_1}(z,C_{46}) = 
    \smashoperator{\int\limits_0^{z}} dF_{T_{11}}(t)
    \smashoperator{\int\limits_0^{t}} dF_{Y_{1}}(y) 
    \smashoperator{\int\limits_0^{z-t}} dF_{S}(x) \smashoperator{\int\limits_{0}^{x}} dF_{T_{12}}(u).$}
\end{align}
\begin{align}
\resizebox{0.89\linewidth}{!}{$
    F_{X_1}(z,C_{46}) = \smashoperator{\int\limits_0^{\infty}} dF_{Y_{1}}(y)
    \smashoperator{\int\limits_y^{y+z}} dF_{T_{11}}(t) 
    \smashoperator{\int\limits_0^{z+y-t}} dF_{S}(x) 
    \smashoperator{\int\limits_{0}^{x}} dF_{T_{12}}(u).$}
\end{align}
The LSTs of $D_{1,i}$ and $X_{1,i}$ give:
\begin{align}
    \eta_1(s,C_{46}) = \rho_{11}\tau_{11}(s)\beta^2(s)\tau_{12}(s+\mu).
\end{align}
\begin{align}
    \xi_1(s,C_{46}) = \rho_{11}\tau_{11}(s)\beta_{1}(s)\tau_{12}(s+\mu).
\end{align}

\paragraph{Case C5} the PDFs of the intervals $D_{1,i}$ and $X_{1,i}$ in the case C5 can be calculated as
\begin{align}
\resizebox{0.89\linewidth}{!}{$
    F_{D_{1,i}}(z,C_5) = 
    \smashoperator{\int\limits_0^{z}}dF_{T_{11}}(t) \smashoperator{\int\limits_0^{t}}dF_{Y_1}(y) \smashoperator{\int\limits_0^{z-t}}dF_{T_{12}}(u) \smashoperator{\int\limits_{0}^{u}}dF_S(x),$}
\end{align}
\begin{align}
\resizebox{0.89\linewidth}{!}{$
    F_{X_{1,i}}(z,C_5) = 
    \smashoperator{\int\limits_0^{\infty}}dF_{Y_1}(y) \smashoperator{\int\limits_y^{y+z}}dF_{T_{11}}(t) 
    \smashoperator{\int\limits_0^{z+y-t}}dF_{T_{12}}(u) \smashoperator{\int\limits_{0}^{u}}dF_S(x).$}
\end{align}
The LSTs $\eta_1(s,A5)$ and $\xi_1(s,A5)$ in the  case C5 yield:
\begin{align}
    \eta_1(s,C5) = \rho_{11}\tau_{11}(s)\beta_{1}(s)(\tau_{12}(s) - \tau_{12}(s+\mu)),
\end{align}
\begin{align}
    \xi_1(s,C5) = \rho_{11}\tau_{11}(s)(\tau_{12}(s) - \tau_{12}(s+\mu)).
\end{align}

The resulting LST of the PAoI distribution of priority packets yields:
\begin{align}
\label{eq:PAoI1}
    \alpha_1(s) =& \Big[\frac{\lambda_1 \nu}{\lambda_1 + s} \beta_{1}(s)\tau_{12}(\lambda_1+s) - \frac{s}{s+\theta}\rho_{11}\beta_{1}(s)\times \nonumber\\
    &\times (\tau_{12}(s) - \tau_{12}(s+\mu)(1 - \nu\beta_{1}(s)))\Big] \beta_1(s),
\end{align}
where $\nu = 1-\rho_2 + \rho_2 \tilde{\beta}_2(s)$. 

The LST of system delay is given as follows:
\begin{align}
\label{eq:Delay1}
    \tau_1(s) =& \Big[ \tau_{11}(s)\tau_{12}(\lambda_1)\Big(\nu - \frac{\lambda_1}{\lambda_1 - s}\Big) + \tau_{12}(s) \times \nonumber\\
    &\times \Big((1-\rho_{11})\frac{\lambda_1}{\lambda_1 - s} + \rho_{11}\tau_{11}(s)\Big)\Big]\beta_1(s).
\end{align}
Given \eqref{eq:AoI} and \eqref{eq:PAoI1}--\eqref{eq:Delay1} the LST of $\Delta_1$ yields:
\begin{align}
\label{eq:AoI1}
    &\delta_1(s) = \beta_1(s) \Big[
    \tau_{11}(s)\tau_{12}(\lambda_1)\frac{\lambda_1}{s - \lambda_1}(\nu + \frac{\lambda_1}{s}(1-\nu)) +  \nonumber\\
    &+ \frac{\lambda_1}{\lambda_1 + s}\beta_{1}(s) \tau_{12}(s)(1+\frac{\lambda_1}{s}(1-\nu)) - \frac{\rho_{11}^3\beta(s)}{1-\rho_{11}}\tau_{11}(s) \times \nonumber\\
    &\times \tau_{12}(\mu+s)\tau_{11}(s) (1-\nu\beta(s)) + \nu\rho_{11}^2\tilde{\beta}(s)\beta(s)
    \Big].
\end{align}

Having the LSTs \eqref{eq:PAoI1}--\eqref{eq:AoI1}, we can calculate the average system delay, PAoI and AoI as $\mathbb{E}[T_1] = - \tau_{1}'(0)$, $\mathbb{E}[A_1] = -\alpha_1'(0)$, and $\mathbb{E}[\Delta_1] = -\delta_1'(0)$: 
\begin{align}
    \mathbb{E}[T_1] = b + \frac{\lambda_1 b^{(2)}}{2(1-\rho_1)} + b_1 + \frac{\lambda_1 b_1^{(2)} + \lambda_2 b_2^{(2)}}{2(1-\rho_1)},
\end{align}
where $b_j^{(k)}$ denote the $k$-th moments of packet $j$ service time.
\begin{align}
    \mathbb{E}[A_1] &= \Big(\frac{1}{\lambda_1} + b_1 + \rho_2 \tilde{b}_2 \Big)\tau_{12}(\lambda_1) - \rho_{11}(b + b\tau_{12}(\mu)) \times \nonumber\\
    &\times (1-\rho_2 + \rho_2 \tilde{b}_2) + (1 - \rho_{11})\Big( b_1 + \mathbb{E}[T_{12}]\tau_{12}(\mu)\Big) + \nonumber\\
    &+ \rho_{11} (1-\rho_2 + \rho_2 \tilde{b}_2)\Big(b_1 + \mathbb{E}[T_{11}] + \mathbb{E}[T_{12}]\tau_{12}(\mu)\Big) + \nonumber \\
    &+ \rho_{11}(b_1 + \mathbb{E}[T_{11}] + \mathbb{E}[T_{12}](1 - \tau_{12}(\mu))).
\end{align}
where $\tilde{b}_2 = b_2^{(2)}/2b_2$ is the average residual service time of non-priority packets.

We give lower bound $\mathbb{E}[\underline{\Delta}_1]$ for the average AoI:
\begin{align}
    &\mathbb{E}[\underline{\Delta}_1]  = b_1 + \frac{1}{\lambda_1} \tau_{12}(\lambda_1) + \tau_{12}(\lambda_1) \mathbb{E}[T_1] +  \rho_1^2 \mathbb{E}[T_1] + \nonumber\\
    & + \rho_{11}^2 \Big(\frac{1}{\theta} - \frac{\rho_{11}}{\mu} + \frac{\rho_{11}}{\theta} + \frac{1}{\lambda_1} + \frac{\mu}{\lambda_1^2} - \frac{1}{\rho_{11}^2} - \frac{1}{\rho_{11}}\Big).
\end{align}

\subsection{Non-priority packets}
\label{sec:NonPriority}

Non-priority packet $i$ can start service only if the second node is free of priority packets, i.e. at the end of the busy period $G_{2,i-1}$ or $G_1$, or if the node is empty. Let us introduce the interval $\Psi_{2,i-1} = W_{2,i-1} + G_{2,i-1}$, where $W_{2,i-1}$ stands for the waiting time of non-priority packet $i-1$. Intervals $W_{2,i-1}$ and $G_{2,i-1}$ are independent, therefore the LST of $\Psi_{2,i-1}$ can be given as $\psi_2(s) = \omega_2(s) \beta_2(s + \lambda_1 - \lambda_1 \gamma(s))$. We consider three cases to define the PAoI $A_{2,i}$. 
\begin{itemize}
    \item [B1:] if $Y_{2,i} > \Psi_{2,i-1}$ and packet $i$ finds the second node empty it immediately goes to service, therefore $A_{2,i} = Y_{2,i} + S_{2,i}$. At the end of interval $\Psi_{2,i-1}$ the node is empty, therefore the probability that packet $i$ finds the node empty upon arrival equals to $1-\rho_1$.
    \item [B2:] if $Y_{2,i} > \Psi_{2,i-1}$ and packet $i$ finds the node busy with a priority packet with probability $\rho_1$ it waits until the end of the ongoing busy period $G_1$, thus $A_{2,i} = Y_{2,i} + \tilde{G}_{1} + S_{2,i}$, where $\tilde{G}_{1}$ denotes the residual time of interval $G_1$. 
    \item [B3:] if $Y_{2,i} < \Psi_{2,i-1}$ packet $i$ finds the second node busy with non-priority packet $i-1$, therefore $A_{2,i} = \Psi_{2,i-1} + S_{2,i}$. 
\end{itemize}

The LST $\alpha_2(s)$ can be given as the sum of three LSTs namely $\alpha_{2}(s,B_1)$, $\alpha_{2}(s,B_2)$ and $\alpha_{2}(s,B_3)$ defined above.   

\paragraph{Case B1} 
the LST of $Y_{2,i} + S_{2,i}$ if $Y_{2,i} > \Psi_{2,i-1}$ and the node is free of priority packets can be given as
\begin{align}
    \alpha_2(s,B_1) 
    = (1-\rho_1)\frac{\lambda_2}{\lambda_2 + s} \psi_2(s+\lambda_2)\beta_2(s).
\end{align}

\paragraph{Case B2} 
the LST of $Y_{2,i} + \tilde{G}_{1} + S_{2,i}$ when $Y_{2,i} > \Psi_{2,i-1}$ and packet $i$ arrives during the busy period $G_1$ takes
\begin{align}
    \alpha_2(s,B_2) = \rho_1\frac{\lambda_2}{\lambda_2 + s}\psi_2(s+\lambda_2)\tilde{\gamma}(s)\beta_2(s),
\end{align}
where $\tilde{\gamma}(s) = (1 - \gamma(s))/\mathbb{E}[G_1]s$ stands for the distribution of the  residual time of the interval $G_1$.

\paragraph{Case B3} if $Y_{2,i} < \Psi_{2,i-1}$ the LST of the PAoI yields
\begin{align}
    \alpha_2(s,B_3) 
    = (\psi_2(s) - \psi_2(s+\lambda_2))\beta_2(s).
\end{align}

The resulting LST of the PAoI distribution of non-priority packets gives
\begin{align}
\label{eq:PAoI2}
    \alpha_{2}(s) =& \Big[ (1-\rho_1)\frac{\lambda_2}{\lambda_2 + s} \psi_2(s+\lambda_2) + \psi_2(s) - \psi_2(s+\lambda_2) + \nonumber\\
    &+ \rho_1 \frac{\lambda_2}{\lambda_2 + s} \psi_2(s+\lambda_2)\tilde{\gamma}(s) \Big] \beta_2(s).
\end{align}
Having \eqref{eq:AoI}, \eqref{eq:SystemDelay2} and \eqref{eq:PAoI2} we give the LST of the AoI distribution of non-priority packets as follows:
\begin{align}
    &\delta_2(s) = \frac{\rho_2}{1 - \rho_1}\tau_2(s)\tilde{\beta}_2(s+\lambda_1 - \lambda_1 \gamma(s)) +\nonumber\\
    &+ \psi(\lambda_2 + s)\beta_2(s) \Big(\frac{\lambda_2}{\lambda + s} + \rho_1 \frac{\lambda_2}{\lambda + s} \frac{\lambda_2}{s} (1 - \tilde{\gamma}(s))\Big),
\end{align}
where $\tilde{\beta}_2(s+\lambda_1 - \lambda_1 \gamma(s))$ denotes the residual time of the busy period $G_2$ and equals to $(1 - \beta_2(s+\lambda_1 - \lambda_1 \gamma(s)))/s\mathbb{E}[G_2]$.

The straightforward calculation of $\alpha_2'(0)$ and $\delta_2'(0)$ gives the average PAoI $\mathbb{E}[A_2]$ and AoI $\mathbb{E}[\Delta_2]$:
\begin{align}
\label{eq:APAoI2}
    \mathbb{E}[A_2] =&  b_2 + \frac{\lambda_1 b_1^{(2)} + \lambda_2 b_2^{(2)}}{2(1-\rho)(1-\rho_1)} +  \frac{b_1}{1-\rho_1} + \psi(\lambda_2) \frac{1}{\lambda_2} + \nonumber\\
    &+ \rho_1 \psi(\lambda_2)\frac{b_2}{2(1-\rho_1)^2}.
\end{align}
\begin{align}
    \mathbb{E}[\Delta_2] =&  \frac{\rho_2}{1-\rho_1} \Big(b_2 + \frac{\lambda_1 b_1^{(2)} + \lambda_2 b_2^{(2)}}{2(1-\rho)(1-\rho_1)} +  \frac{b_1}{1-\rho_1} \Big) + \nonumber\\
    &+ \psi(\lambda_2)\Big(\frac{1}{\lambda_2} + \frac{\rho_1\lambda_2}{2} \Big(\frac{b_1^{(3)}/b_1^{(2)}}{3(1-\rho_1)} + \frac{\lambda_1 b_1^{(2)}}{(1-\rho_1)^2} \Big)\Big) + \nonumber\\
    &+ \psi(\lambda_2)\Big(1 + \frac{\rho_1 \rho_2}{2(1-\rho_1)^2}\Big)(b_2 + \psi'(\lambda_2)).
\end{align}

\section{Selected numerical results}
\label{sec:Results}

The results of our analysis have been validated by Monte Carlo simulation. All data collected during the transient state has been discarded. We model  arrivals, service and departures of $10^5$ packets of the reference system. 
We calculate the average PAoI, AoI and system delay for different values of $p = \{0.1, 0.3, 0.5, 0.7, 0.9\}$ to capture the effect of the status updates generation rate on the AoI. The numerical results are given under the assumption of exponential service time with means $b = b_1 = b_2 = 1$ for variable utilization $\rho = \{0.1,\dots, 0.9\}$ at the second node.

\begin{figure*}[bht]
    \centering
    \label{fig:Priority}
    \subfigure[Average PAoI]{\includegraphics[scale=0.29]{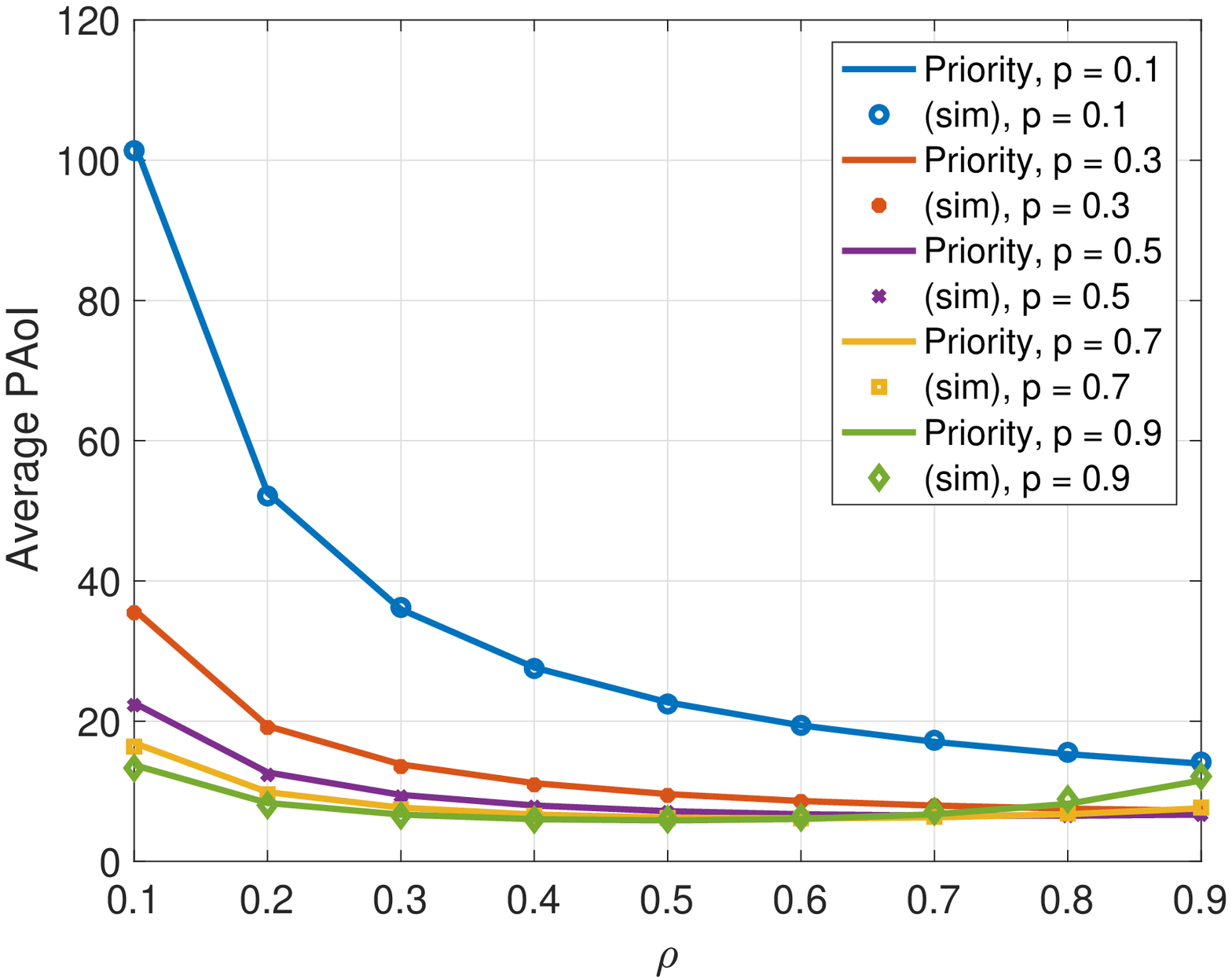}
    	\label{fig:Priority1}}
    \subfigure[Average AoI]{\includegraphics[scale=0.29]{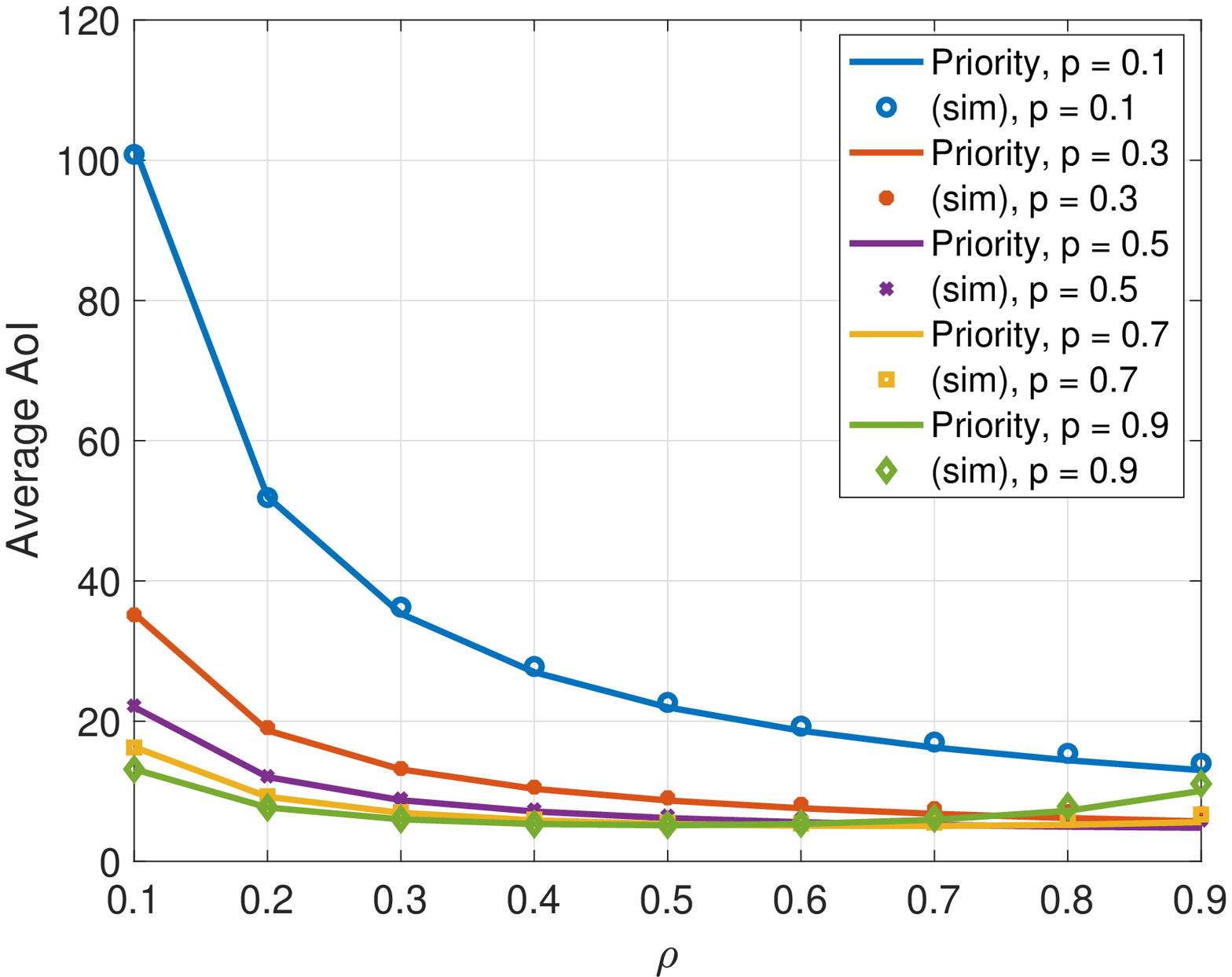}
    	\label{fig:Priority2}}
    \subfigure[Average system delay]{\includegraphics[scale=0.29]{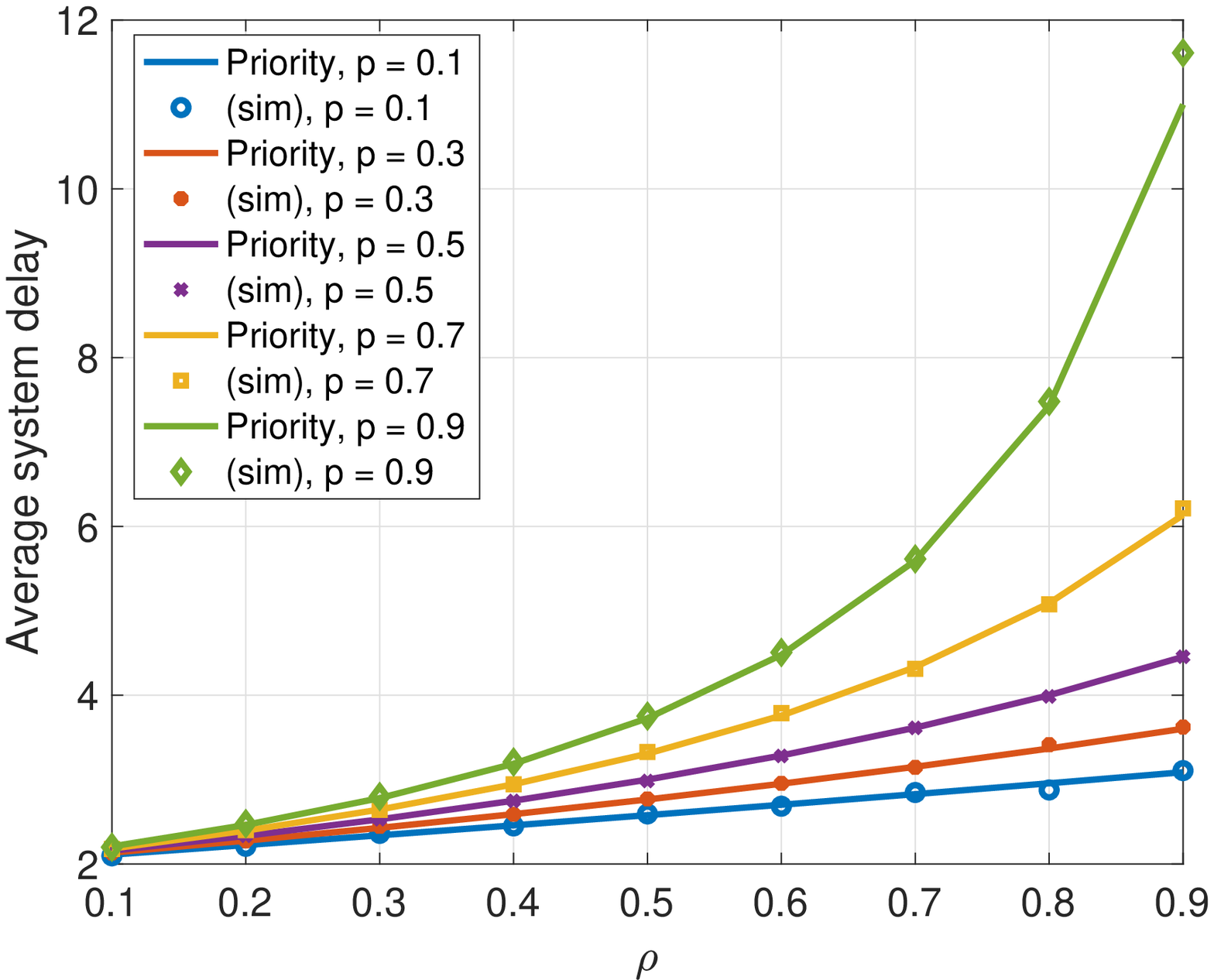}
    	\label{fig:Priority3}}
    \vspace{-0.2cm}
    \caption{Average PAoI (a), AoI (b) and system delay (c) of priority packets}
\end{figure*}

\begin{figure*}[hbt]
    \centering
    \label{fig:NonPriority}
    \subfigure[Average PAoI]{\includegraphics[scale=0.29]{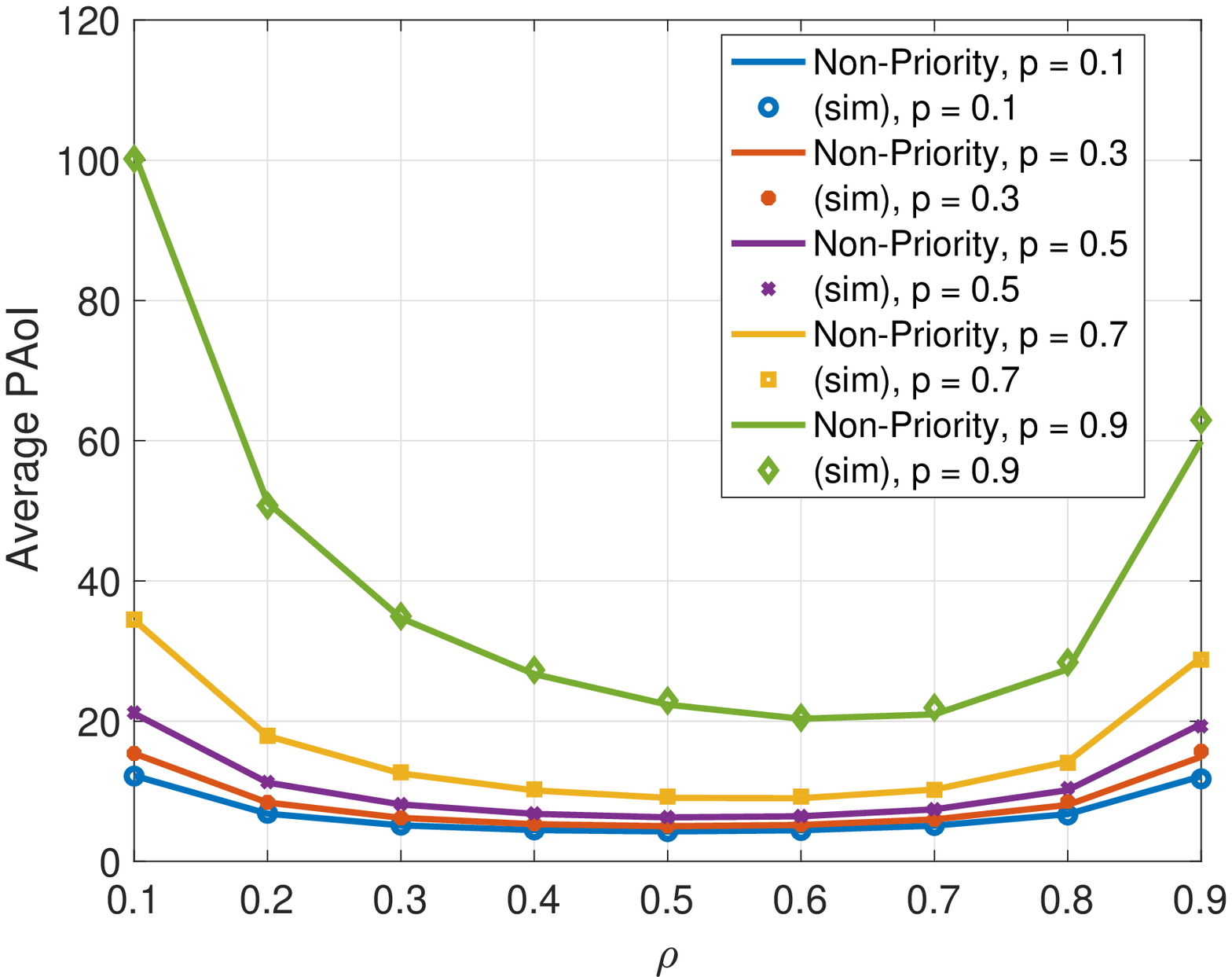}
    	\label{fig:NonPriority1}}
    \subfigure[Average AoI]{\includegraphics[scale=0.29]{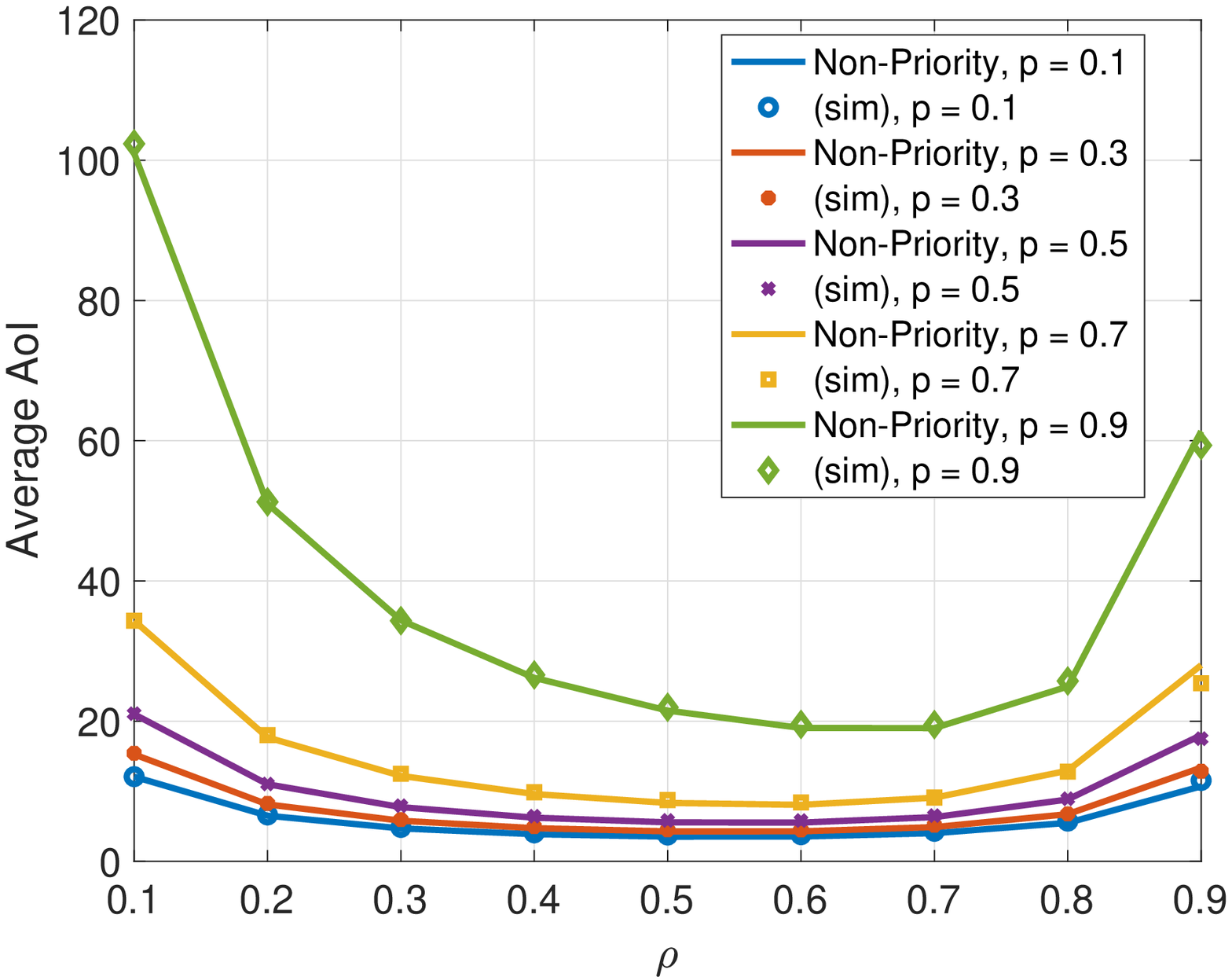}
    	\label{fig:NonPriority2}}
    \subfigure[Average system delay]{\includegraphics[scale=0.29]{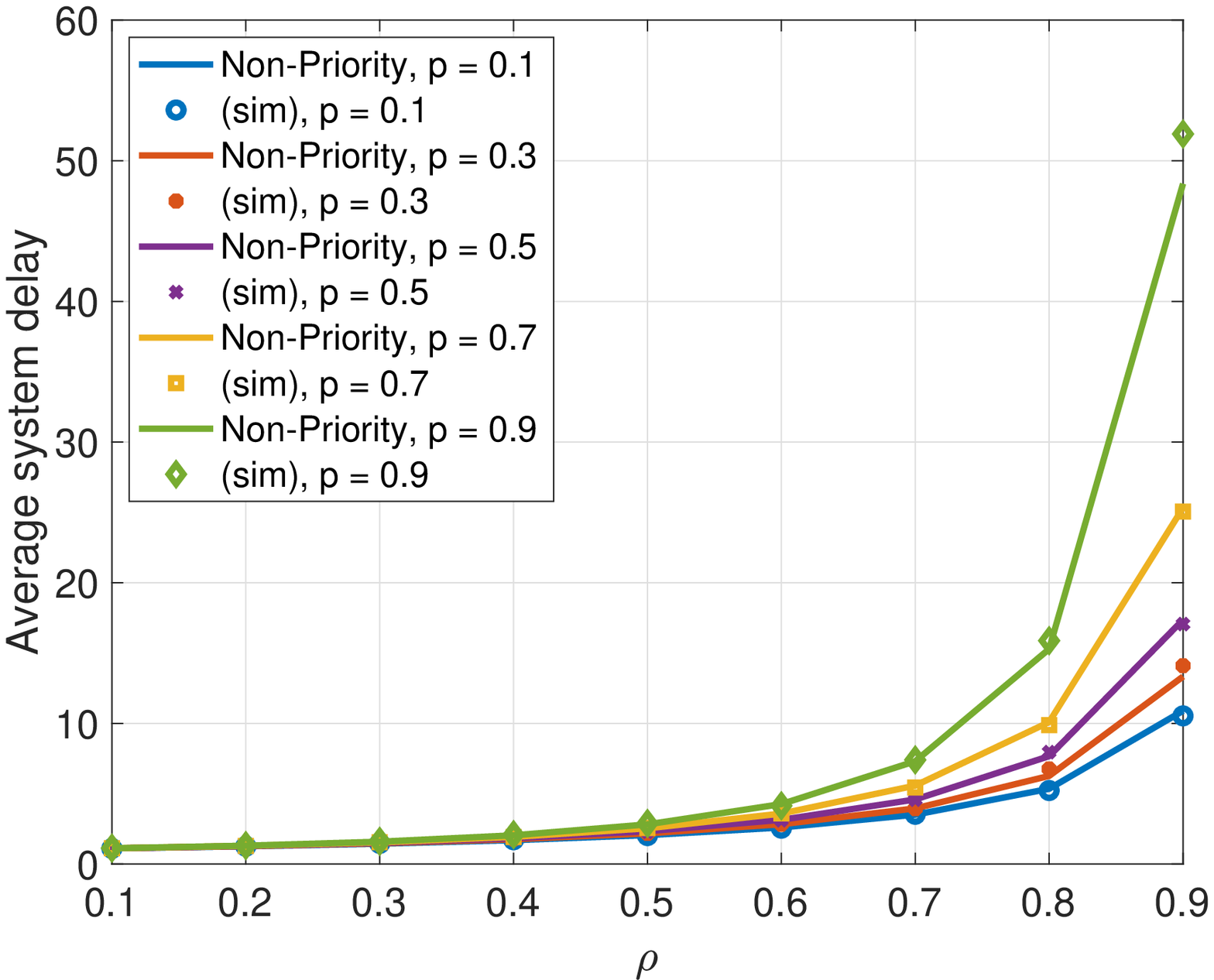}
    	\label{fig:NonPriority3}}
    \vspace{-0.2cm}
    \caption{Average PAoI (a), AoI (b) and system delay (c) of non-priority packets}
\end{figure*}

The metrics of interest of priority packets are depicted in Fig. 3. 
The simulation results of the PAoI illustrated in Fig.~\ref{fig:Priority1} show a perfect fit of our bound with the analytical curves, which justifies the assumption that priority packet $i$ finds non-priority packets at the second node with the given probability. 
The results for non-priority packets in Fig. 4 are instead exact. The given lower bound for AoI is tight when the system utilization is low and becomes more visible when $\rho$ increases. In our system, the PAoI is a tight upper bound of the AoI due to the low correlation between interarrival and delay intervals of consecutive packets.
The average AoI of priority packets decreases when the status update rate increases if the priority system utilization $\rho_1 < 0.63$. If $\rho_1 \geq 0.63$ the AoI gradually increases demonstrating a wide U shape, the AoI of non-priority packets shows similar results in Fig.~\ref{fig:NonPriority2}. This means that the optimal performance can be reached. 

Besides the average AoI the average PAoI and system delay of non-priority packets are shown in Fig.~\ref{fig:NonPriority1} and Fig.~\ref{fig:NonPriority3} respectively. Again the average PAoI is a very tight upper bound for the AoI. Due to the non-priority packets preemption in waiting the average system delay rapidly increases when the utilization at the second node increases. Both PAoI and AoI of non-priority packets depend on the system delay more than that of priority packets. If non-priority packets may tolerate a certain packet error rate also due to the discarding of outdated packets the AoI could be improved if a newly arrived non-priority packet replaces the previously queued packet. 


\section{Conclusions}
\label{sec:Conclusions}
In this paper we have investigated the timeliness of the status updates in a multi-hop IoT tracking system with two nodes and different entry points for priority and non-priority traffic. We have derived the distribution of AoI, PAoI and system delay in terms of LST and have given closed-form expressions for their first moments. We have obtained the exact expressions for non-priority packets and tight bounds for priority flow of packets. In our system, PAoI is a tight upper bound for both classes of traffic.

The extension to $N$ hops requires an exponential service time at first $N-1$ hops while the last hop that aggregates traffic from all previous hops holds general service time distribution. Such an assumption is in line with many multi-hop systems from the reference literature.
Other possible research directions are the extension to more priority levels, LCFS discipline with packets discarding, and age-aware packet management. 


\bibliographystyle{IEEEtran}
\bibliography{Globecom_AOI}

\end{document}